\documentclass{article}
\usepackage{ijcai24}

\usepackage{times}
\usepackage{soul}
\usepackage{url}
\usepackage[hidelinks]{hyperref}
\usepackage[utf8]{inputenc}
\usepackage[small]{caption}
\usepackage{graphicx}
\usepackage{amsmath}
\usepackage{amsthm}
\usepackage{booktabs}
\usepackage{algorithm}
\usepackage{algorithmic}
\usepackage[switch]{lineno}

\usepackage[table,xcdraw]{xcolor}
\usepackage{graphicx} 
\usepackage{multirow}
\usepackage[numbers,sort&compress]{natbib}

\usepackage{algorithmic}
\usepackage{algorithm}

\title{Are You Copying My Prompt? Protecting the Copyright of Vision Prompt for VPaaS via Watermark}

\author{	Huali Ren$^{1}$, Anli Yan$^2$, Chong-zhi Gao$^{3*}$, Hongyang Yan$^2$, Zhenxin Zhang$^4$, Jin Li$^2$\\
	\affiliations
	$^1$School of Cyberspace Security, Guangzhou University, China\\
	$^2$Institute of Artificial Intelligence, Guangzhou University, China\\
	$^3$School of Computer Science, Guangzhou University, China\\
	$^4$School of Cyber Engineering, Xidian University, China \\
	\emails
	czgao@gzhu.edu.cn
}

\begin{document}
	
	\maketitle
	
	\begin{abstract}
		
	Visual Prompt Learning (VPL) differs from traditional fine-tuning methods in reducing significant resource consumption by avoiding updating pre-trained model parameters. Instead, it focuses on learning an input perturbation, a visual prompt, added to downstream task data for making predictions. Since learning generalizable prompts requires expert design and creation, which is technically demanding and time-consuming in the optimization process, developers of Visual Prompts as a Service (VPaaS) have emerged. These developers profit by providing well-crafted prompts to authorized customers. However, a significant drawback is that prompts can be easily copied and redistributed, threatening the intellectual property of VPaaS developers. Hence, there is an urgent need for technology to protect the rights of VPaaS developers. To this end, we present a method named \textbf{WVPrompt} that employs visual prompt watermarking in a black-box way. WVPrompt consists of two parts: prompt watermarking and prompt verification. Specifically, it utilizes a poison-only backdoor attack method to embed a watermark into the prompt and then employs a hypothesis-testing approach for remote verification of prompt ownership. Extensive experiments have been conducted on three well-known benchmark datasets using three popular pre-trained models: RN50, BIT-M, and Instagram. The experimental results demonstrate that WVPrompt is efficient, harmless, and robust to various adversarial operations.

	\end{abstract}
	
	\begin{figure*}[t]
		\centering
		\includegraphics[width=0.95\textwidth]{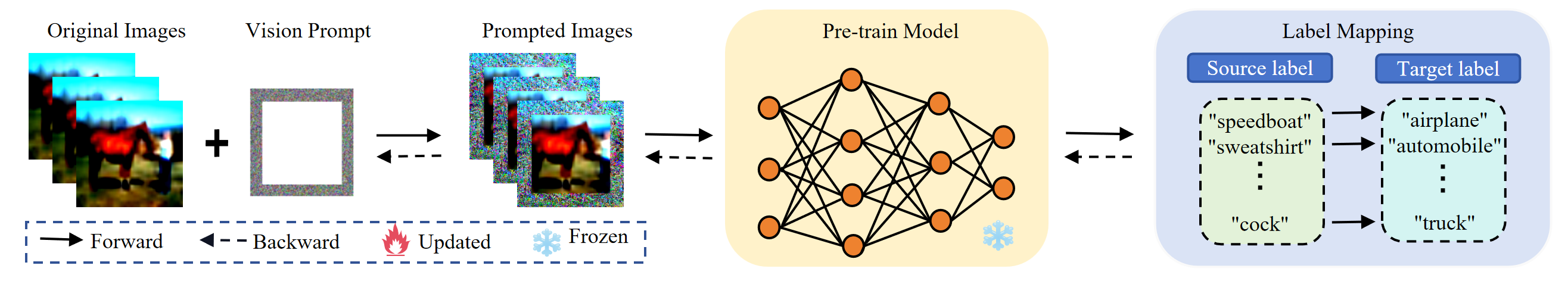} 
		\caption{The workflow of visual prompt learning}
		\label{visionprompt}
	\end{figure*}
	
	\section{Introduction}
	
	The 'pre-training + fine-tuning' paradigm has demonstrated considerable success in applying large pre-trained models to various downstream tasks \cite{zhuang2020comprehensive}. However, fine-tuning often incurs significant computational overhead due to the need to record gradients for all parameters and the state of the optimizer. Additionally, the pre-trained model post-fine-tuning becomes task-specific, leading to substantial storage costs for maintaining separate copies of the model's backbone parameters for each downstream task \cite{bahng2022exploring,elsayed2018adversarial}.

	Inspired by recent advancements in Natural Language Processing (NLP) prompts \cite{li2021prefix,lester2021power,shin2020autoprompt}, Visual Prompt Learning (VPL) \cite{bahng2022exploring,huang2023diversity,bar2022visual,chen2023understanding} emerges as a promising solution to address several pertinent challenges. Unlike traditional fine-tuning, VPL employs input and output transformations to adapt a fixed pre-trained model for diverse downstream tasks (see Figure \ref{visionprompt}). The input transformation involves learning input perturbations (prompts) to transform data from downstream tasks into the original data distribution through padding or patching. On the other hand, the output transformation is achieved through a label mapping (LM) function, which maps source labels to target labels \cite{bahng2022exploring}. For instance, in the CIFAR10 classification task, the fine-tuned pre-trained model RN50 contains 25 million parameters, whereas the prompt with a padding size P of 30 only has 60,000 parameters.
	
	However, designing and optimizing appropriate prompts for Visual Prompt Learning (VPL) remains a challenge that demands substantial effort. This challenge has led to the emergence of Visual Prompts as a Service (VPaaS), aimed at facilitating VPL applications for non-expert users. VPaaS developers require abundant data, expertise, and computing resources to optimize prompts, thus making prompts a valuable asset. Customers can purchase and utilize these prompts with pre-trained vision models for making predictions. Despite the convenience of visual prompts, their susceptibility to unauthorized replication and redistribution poses significant threats to VPaaS developers' interests \cite{shen2023prompt,li2023feasibility}. Furthermore, unauthorized prompts can serve as a springboard for malicious attackers, resulting in the exposure of VPaaS developer's privacy. Recent studies, such as \cite{backesquantifying}, have underscored the risks associated with using attribute inference attacks and member inference attacks to extract privacy from visual prompts. This vulnerability raises concerns about safeguarding intellectual property (IP) rights linked to visual prompts, an issue demanding urgent attention.
	
	In the current field of artificial intelligence, IP protection mainly focuses on safeguarding AI models and datasets. The primary defense methods include fingerprinting \cite{cao2021ipguard,pan2022metav,chen2022copy,yang2022metafinger,guan2022you}, watermarking \cite{yao2023promptcare,adi2018turning,lukas2022sok,li2023black,li2022untargeted, shafieinejad2021robustness}, and dataset inference \cite{maini2021dataset,dziedzic2022dataset}. Among these techniques, watermarking stands out as a classic and intuitive approach with the greatest potential for visual prompt copyright protection. Typically, watermarking \cite{adi2018turning} leverages the over-parameterization of models to embed identity information without compromising model utility, and then verifies ownership by extracting and comparing watermark information. Additionally, watermarking technology has recently been applied to large-scale model content generation detection \cite{kirchenbauer2023watermark,wang2023towards}, aiding in distinguishing between human-written and machine-generated content. However, compared to models, the parameters in visual prompts are significantly reduced, leading us to naturally believe that this paradigm has lost the characteristic of over-parameterization, making it difficult to effectively embed watermarks. Additionally, it is questionable whether embedded watermarks can withstand some post-processing operations that remove watermarks, such as prompt fine-tuning and pruning. To the best of our knowledge, there is no prior work to solve it.

	In this paper, we formulate the IP protection of visual prompts as an ownership verification problem, where a VPaaS developer (also referred to as a defender) seeks to determine whether a suspicious prompt is an unauthorized reproduction of its prompt. Compared to the white-box setting, we are focusing on the more challenging and practical black-box setting. In this scenario, the defender can only access prediction results from the pre-trained model vendor through an API, without any training details or parameter information about the suspicious prompts.
	To address these challenges, we propose a backdoor-based prompt watermarking method called WVPrompt. WVPrompt consists of two main steps: prompt watermarking and prompt verification. Specifically, we employ a pure poison backdoor attack \cite{gu2019badnets,li2021invisible,nguyen2021wanet} for prompt watermarking. The prompt maintains high prediction accuracy on clean samples while exhibiting specific behavior on samples with special triggers. Simultaneously, the defender verifies whether the suspicious prompt is a pirated copy of the target prompt by checking for the presence of a specific backdoor, completing the watermark verification process. To achieve this, we design a validation method guided by hypothesis testing. A series of experiments demonstrate the effectiveness of WVPrompt.

	Our contributions are as follows:
	\begin{itemize}
	\item We conducted the first systematic investigation of IP protection in VPaaS, exploring the risks of unauthorized prompt utilization.
	\item We designed a black-box prompt ownership verification system (WVPrompt), leveraging poison-only backdoor attacks and hypothesis testing.
	\item We conducted comprehensive experiments on three renowned benchmark datasets, using three popular visual prompt learning methods to evaluate three pre-trained models (RN50, BIT-M, Instagram). It proves the effectiveness, harmlessness and robustness of WVPrompt.
	\end{itemize}
	
	\section{Preliminaries}
	
	\subsection{Vision Prompt Learning}
	Visual Prompt Learning (VPL) \cite{bahng2022exploring,huang2023diversity,bar2022visual,chen2023understanding} is a novel learning paradigm proposed to address the computational and storage constraints associated with fine-tuning large models for downstream tasks adaptation. Its primary objective is to acquire a task-specific visual prompt, enabling the reuse of pre-trained models across diverse tasks. Domain experts design an appropriate visual prompt, refined through extensive data analysis and computational resources. Generating visual prompts involves two key stages: input and output transformations.
	
	\textbf{Input transformation}. This phase focuses on devising a suitable prompt to convert downstream target task data $D_{t}$ into the source data $D_{s}$ distribution. To achieve this, visual prompts inject a task-relevant perturbation pattern $\delta$ into the pixel space of the input sample $x$, where $x\in R^{N_{t}}$. The general form is expressed as:

	\begin{equation}
		x^{'} (\delta )=h(\delta,x)\in R^{N_{s} } , x\in R^{N_{t} }
	\end{equation}
	Here, $h(.,.)$ denotes the input conversion function, with perturbation patterns typically encompassing random position patches, fixed position patches, and padding \cite{bahng2022exploring}. Most studies utilize additive transformation functions based on padding patterns, as illustrated in Figure \ref{visionprompt}.

	Once the prompt model in Equation (1) is designed and parameters initialized, given a frozen target pre-training model $f_{t}$ and a downstream task dataset $D_{t}=\left \{ (x_{1},y_{1}),...,(x_{m},y_{m}) \right \} $, the training process resembles that of generating high-precision models under supervised learning. The cross-entropy loss function $L(.,.)$ guides backpropagation and the stochastic gradient descent method optimizes the prompts. The specific losses are defined as:
	\begin{equation}
		\underset{\delta }{arg min} E_{(x,y)\in D_{t} }[L(f_{t}(x^{'} (\delta )),y)] 
	\end{equation}
	
	\textbf{Output transformation}: In classification tasks, the output category $K_{s}$ of the source pre-trained model often differs from the target downstream task category $K_{t}$, typically $K_{t} < K_{s}$. Thus, establishing a suitable label mapper is necessary to create a one-to-one correspondence between the source label space $Y_{s}$ and the target label space $Y_{t}$, facilitating the direct prediction of the correct target label by the pre-trained model. Current research primarily explores three tag mapping methods: \textcircled{1} random mapping \cite{bahng2022exploring,elsayed2018adversarial}, \textcircled{2} pre-defined, one-shot frequency-based mapping \cite{tsai2020transfer,chen2021adversarial}, and \textcircled{3} iterative frequency-based mapping \cite{chen2023understanding}. These methods are discussed in detail below.
	
	\begin{figure}[t]
		\centering
		\includegraphics[width=0.35\textwidth]{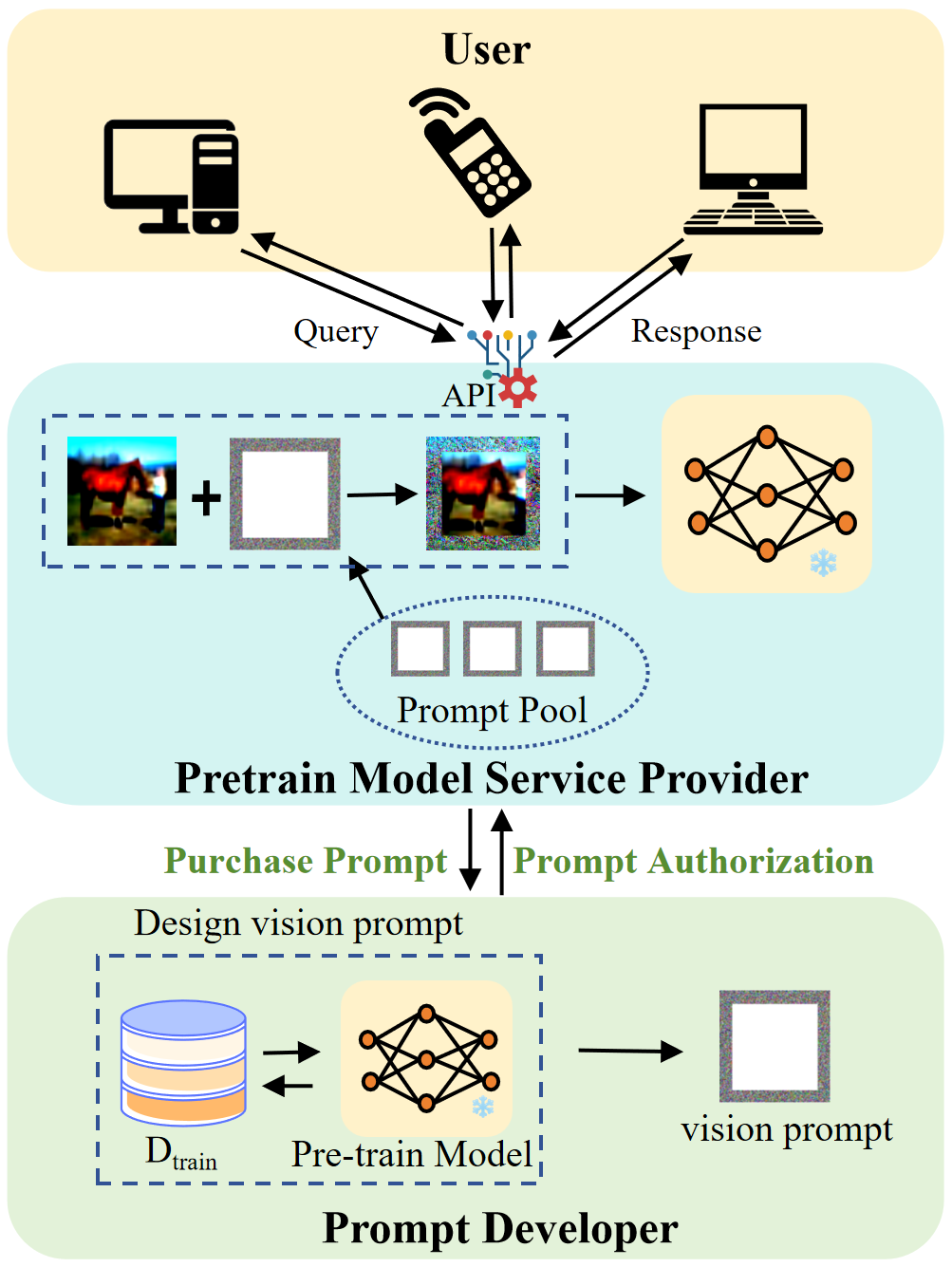} 
		\caption{Illustration of Vision Prompt as a Service (VPaaS) pipeline}
		\label{prompt_server}
	\end{figure}
	
	\textcircled{1}\textbf{Random Label Mapping-Based Visual Prompting (RLMVP)}: RLMVP does not leverage any prior knowledge of the pre-trained model and the source dataset to guide the label mapping process. It randomly selects $K_{t}$ labels from the source label space $Y_{s}$ and maps them directly to the target labels. For instance, when employing the pre-trained RN50 model on the source task dataset (ImageNet) for the downstream target dataset (SVHN), the top 10 model outputs are typically directly used as target indices, i.e., ImageNet label i → SVHN label i, despite the lack of interpretation.

	\textcircled{2}\textbf{Frequency label mapping-based visual prompting (FLMVP)}, FLMVP selects the top $K_{t}$ labels with the highest output label frequency in $f_{t}$. It utilizes $\delta=0$, inputs $X+\delta$ into $f_{t}$, and maps a target label $K_{t}$ to the source label $K_{s}$ according to the equation:
	\begin{equation}
		\begin{aligned} y_{s} (y_{t} ) = \underset{y_{s}}{argmax}\ Pr\left \{ Top-1\ prediction\ of\ f_{t}(h(0,x))\ \right. \\
			\left. is\ y_{s} \mid \forall x_{t}\in \Gamma_{y_{t}} \right \}\end{aligned}
	\end{equation}
	Here, $y_{s}(y_{t})$ explicitly expresses the dependence of the mapped source label on the target label, $\Gamma_{y_{t}}$ represents the target dataset in class $y_{t}$, and $Pr(.,.)$ is the top-1 prediction of $f_{t}$, representing the probability of the source class under the zero-padded target data point in $\Gamma_{y_{t}}$.
	
	\textcircled{3}\textbf{Iterative Label Mapping-Based Visual Prompting (ILMVP)}: Building upon FLMVP, ILMVP accounts for dynamic changes in the mapping between the source label space and the target label space, incorporating interpretable considerations. It employs a two-layer iterative optimization method (BLO) to optimize the underlying label mapping and prompts. In each epoch, it uses the $\delta$
	(non-zero) from the previous round of optimization and employs the frequency calculation method in FLMVP to map the target label $y_{t}$ to the source label $y_{s}$
	, as shown in Equation (4). Subsequently, Equation (2) is utilized to calculate the loss for backpropagation optimization of the prompt:
	
	\begin{equation}
		\begin{aligned} y_{s} (y_{t} ) = \underset{y_{s}}{argmax}\ Pr\left \{ Top-1\ prediction\ of\ f_{t}(h(\delta,x))\ \right. \\
			\left. is\ y_{s} \mid \forall x_{t}\in \Gamma_{y_{t}} \right \}\end{aligned}
	\end{equation}

	\begin{figure*}[t]
		\centering
		\includegraphics[width=0.8\textwidth]{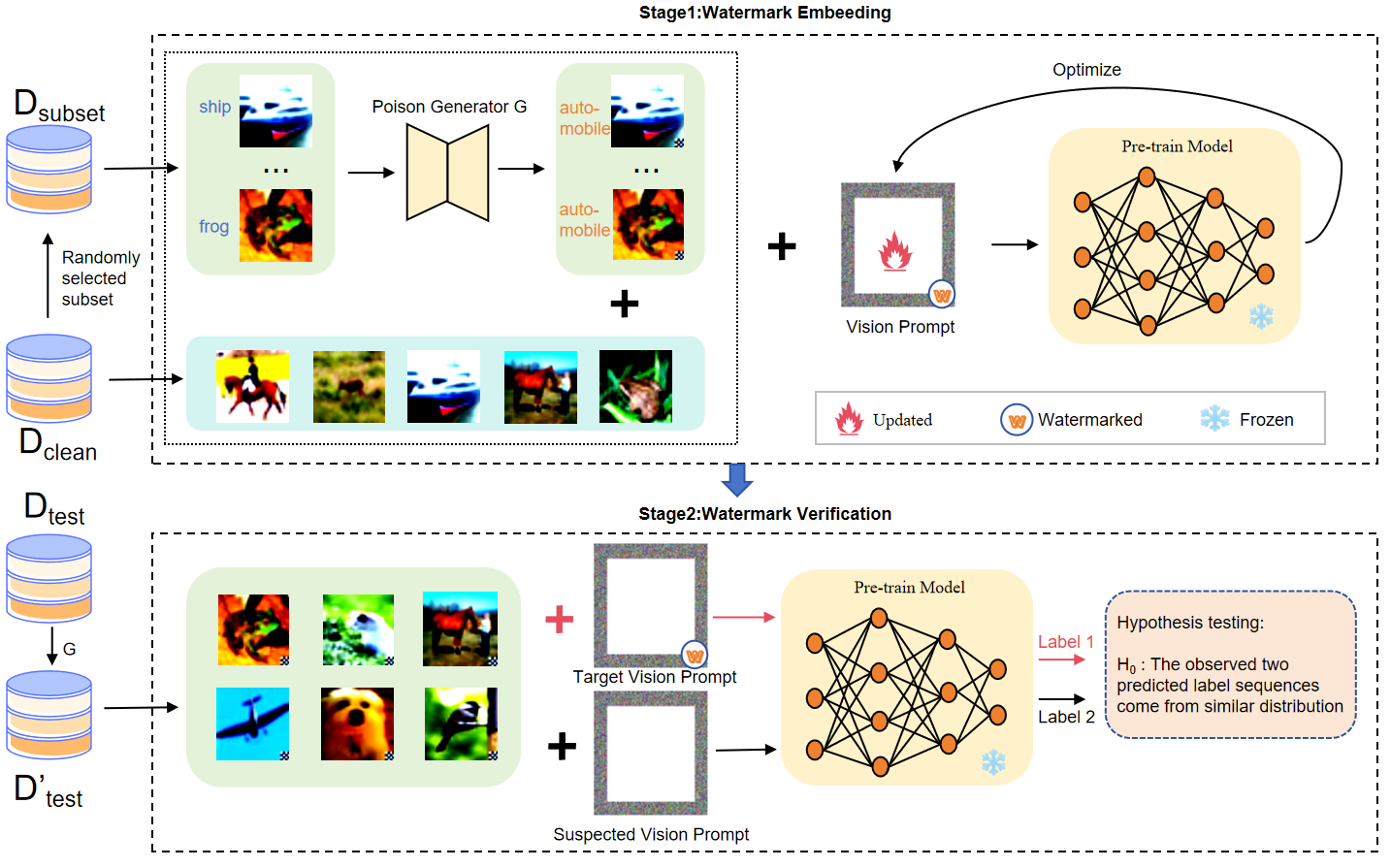} 
		\caption{The pipeline of WVPrompt. In the first step, defenders will exploit poison-only backdoor for prompt watermark embedding. In the second step, the defender will perform prompt ownership verification by checking whether the suspected prompt contains a specific hidden backdoor through hypothesis testing.}
		\label{WVPrompt}
	\end{figure*}
	
	\subsection{Vision Prompt-as-a-Service}
	Visual Prompting as a Service (VPaaS) has seen rapid development, as illustrated in the Praas pipeline in Figure \ref{prompt_server}. Like prompt learning in large language models \cite{yao2023promptcare}, visual prompting involves three primary stakeholders: VPaaS developers, pre-trained model service providers, and end users. VPaaS developers utilize their downstream task datasets and pre-training model services they have purchased to collaboratively optimize the training of visual prompts based on specific requirements. These crafted prompts, which demand significant data and computational resources, are sold or shared with authorized pre-trained model service providers to build a prompt library. Customers who procure services from a pre-trained model provider submit query requests via the public API. Subsequently, the model provider selects suitable prompts from the library, integrates them with the query content, and delivers the computed results back to the customer. In this business model, unauthorized acquisition of visual prompts by a pre-trained model service provider, such as model theft, malicious copying, or distribution, can lead to substantial economic losses for VPaaS developers and serious infringement of their intellectual property (IP). Therefore, protective technologies must be developed to safeguard the IP of VPaaS developers while allowing for external verification of prompt ownership.

	\subsection{Threat model}

	This paper focuses on protecting the copyright of visual prompts through watermarking. The scenario involves two entities: the attacker and the defender. We assume the defender to be a VPaaS developer aiming to publish its prompts and detect copyright infringements in suspicious prompts. The defender has full control over prompts but can only access the pre-trained model provider's API in a black-box manner to obtain classification output labels. During the training phase, the defender can covertly embed watermark information into prompts and subsequently compare the consistency between the extracted and embedded watermark information during verification to detect piracy. On the other hand, the attacker is a malicious pre-trained model service provider attempting to obtain prompts through unauthorized copying or theft. The attacker refrains from training specialized downstream datasets for task prompts but possesses some understanding of task input and output. Additionally, the attacker can fine-tune and prune pirated prompts to evade piracy detection. Both the attacker and defender have access to the pre-trained model.
	
	\section{Methodology}
	In this section, we will first outline the main pipeline of our approach and then delve into its components in detail. Generally, a watermarked visual prompt should satisfy the following three properties to ensure its usability. Here, $\delta$ represents the watermarked prompts, and $\hat{\delta}$ represents the clean prompts.
	
	\textbf{Effectiveness}: For a sample $x'$ with a trigger, the behavior difference between the watermarked $\hat{\delta }$ and the clean $\delta$ should be relatively large in the pre-trained model so that the false accusations against $\delta$ can be distinguished and reduced. 
	
	\textbf{Harmlessness}: The watermarking algorithm should have a negligible impact on the usefulness of the vision prompt in the target downstream tasks, maintaining its functionality.
	
	\textbf{Robustness}: Watermarked visual prompts should be resistant to common post-processing removals attacks such as prompt fine-turn and prompt pruning, preventing attackers from easily bypassing copyright detection mechanisms. Once embedded, it is difficult for plagiarists to remove it.
	
	\subsection{Overview}
	This paper assumes the defender embeds secret information into $\delta$ to protect prompt intellectual property (IP). As depicted in Figure \ref{WVPrompt}, our method consists of two primary steps: (1) watermark injection and (2) watermark verification. Specifically, we employ a backdoor-based method to inject watermarks into prompts and design hypothesis tests to guide prompt verification. The technical details of each step will be described in the following subsections.

	\subsection{Watermark injection}
	We consider a visual prompt $\delta$ added to a raw image $x$ and then processed by an API within a pre-trained model $f$ to generate the mapped label $y$. The VPaaS developer (the defender) aims to embed secret watermark information into $\delta$, transforming it into a watermarked version $\hat{\delta}$ that ensures effectiveness, harmlessness, and robustness. The defender possesses a comprehensive understanding of the complete training process for $\delta$ and has access to the entire downstream task dataset $D_{c} =\left \{ (x_{i},y_{i})\right \} _{i=1}^{N}(x_{i}\in X, y_{i}\in Y) $, where $x_{i} \in X$ and $y_{i} \in Y$ represent the input and output spaces of the downstream data.

	To embed the watermark, the defender randomly selects a subset $D_{s}$ from $D_{c}$ to generate a modified version $D_{p}$ using a specialized poisoning generator $G$ and a target label $y_{t}$. Typically, $D_{s} \in D_{c} $ and $D_{p}=\left \{ (x',y_{t})\mid x'=G(x),(x,y)\in D_{s}\right \}$. The final poisoned dataset $D_{p}$ is merged with the original clean dataset $D_{c}$ to create the watermark dataset $D$. Particularly, $r_{p} = \frac{\mid  D_{p} \mid }{\mid  D_{c} \mid}$ is recorded as the poisoning ratio. In this paper, we follow the common construction method of the poisoning generator $G$ \cite{gu2019badnets}:

	\begin{equation}
		x' = G(x) = (1-\alpha )\otimes x +\alpha \otimes t
	\end{equation}
	
	Here, $t$ represents the trigger mode, $\alpha$ denotes the trigger's position, and this paper adopts a cross iteration of black and white pixel blocks to construct $t$. Simultaneously, the following loss function performs watermark injection on $\delta$.
	
	\textbf{Effectiveness}. To ensure differentiability, we need to ensure that the watermarked $\hat{\delta}$ and the clean $\delta$ yield different predictions when combined with the triggered image $x$ and inputted into the pre-trained model $f_{t}$. As the original real label of $x$ in $D_{p}$ is not $y_{t}$, and the label after poisoning is $y_{t}$, it is essential to ensure that the sample $x'$ with a trigger minimizes the loss under both $\hat{\delta}$ and $f$. The losses are as follows:
	\begin{equation}
		L_{p} = \frac{1}{\mid D_{p}\mid }\sum_{(x',y_{t})\in D_{p}}^{}L(f(h(\hat{\delta },x' )),y
		_{t}) 
	\end{equation}
	
	\textbf{Harmlessness}. We need to ensure that embedding watermarks will not affect the usefulness of prompt $\hat{\delta}$
	on clean samples $x$, i.e., the prediction accuracy in visual classification tasks. Therefore, we follow the standard prompt training process and ensure that $x$ minimizes the loss with $y_{t}$ under both $\hat{\delta}$ and $f$. The loss is as follows:
	
	\begin{equation}
		L_{c} = \frac{1}{\mid D_{c}\mid }\sum_{(x',y_{t})\in D_{c}}^{}L(f(h(\hat{\delta },x)),y) 
	\end{equation}
	
	Based on the above analysis, the watermark injection process is then formulated as an optimization problem:
	\begin{equation}
		\underset{\hat{\delta } }{argmin}(L_{c}+\beta \cdot L_{p})
	\end{equation}
	
	Here $L(.,.)$ denotes a cross-entropy loss function, $\beta$ represents a hyperparameter used to balance distinguishability and harmlessness, and $\beta$ is one unless otherwise specified. We apply the stochastic gradient optimization method to solve this optimization problem.

	\textbf{Robustness}: In conventional machine learning trained neural network models, backdoor watermarks usually show high Robustness to common model transformations (such as model pruning and fine-tuning). For visual prompts, we believe that the watermarking method in this paper is also robust in prompting fine-tuning and pruning. We verify this conclusion in Section 4.4. 
	
	\subsection{Watermark verification}
	Give a suspicious prompt to $\delta$, prompting the owner (VPaaS developer) to verify whether it originates from $\hat{\delta}$ without IP authorization by detecting a specific watermark backdoor. Let $x'$ represent the poisoned sample and $y_{t}$ denote the target label. By querying the pre-trained model 
	$f$, the owner can check for suspicious prompts via the results of $f(h(\delta,x'))$. If $f(h(\delta,x'))=y_{t}$, the suspicious prompt is regarded as a piracy prompt. However, this result may be affected by the randomness of selecting sample $x'$. Therefore, we introduce a hypothesis testing guided method in this paper to enhance the credibility of the verification results.
	
	\textbf{Proposition 1} (Vision Prompt Ownership Verification), assuming $C(x)$ represents the predicted label obtained by inputting the pre-trained model $f$, such as $C(x)=f(h(\delta ,x))$. Let variable $X$ denote clean samples of non-target label $y_{t}$, with the size of $X$ denoted as $m$. $X'$ is the version generated by the poisoner $G$, such as $X' = G(X)$. Given a full hypothesis $H_{0}:C(X')\ne y_{t},(H_{1}:C(X')= y_{t})$, where $y_{t}$ is the predefined target label. We claim the suspicious prompt $\delta$ is a pirated copy of $\hat{\delta}$  if and only if $H_{0}$ is rejected.
	
	In practice, without specific instructions, we randomly select $m$ $(m \ne 100)$ distinct benign samples with non-target labels. We conduct a one-tailed paired t-test \cite{hogg2013introduction} and compute its p-value. Experimental results represent the average of three random selections. If the p-value is below the significance level $\alpha$, we reject the null hypothesis $H_{0}$.
	
	\begin{table}
		\renewcommand\arraystretch{1.2}
		\caption{Statistics and division of datasets}
		\resizebox{\linewidth}{!}{
		\begin{tabular}{c|c|c|c|c|c}
			\hline
			Dataset & \begin{tabular}[c]{@{}c@{}}Train\\  size\end{tabular} & \begin{tabular}[c]{@{}c@{}}Test \\ size\end{tabular} & \begin{tabular}[c]{@{}c@{}}Validation \\ size\end{tabular} & \begin{tabular}[c]{@{}c@{}}Class\\  numbles\end{tabular} & \begin{tabular}[c]{@{}c@{}}Rescaled \\ resolution\end{tabular} \\ \hline
			Cifar10 & 50000                                                 & 5000                                                 & 5000                                                       & 10                                                       & 32×32                                                          \\ \hline
			EuroSAT & 21600                                                 & 2700                                                 & 2700                                                       & 10                                                       & 64×64                                                          \\ \hline
			SVHN    & 73257                                                 & 13016                                                & 13016                                                      & 10                                                       & 32×32                                                          \\ \hline
		\end{tabular}
		}
		\label{Dataset summary}
	\end{table}
	
	\begin{table}
		\renewcommand\arraystretch{1.2}
		\caption{Overview of pretrained models}
		\resizebox{\linewidth}{!}{
		\begin{tabular}{c|c|c|c}
			\hline
			Model     & Architecture & Pre-train Dataset     & Parameters \\ \hline
			RN50      & ResNet-50    & 1.2M ImageNet-1K      & 25,557,032 \\ \hline
			BiT-M     & ResNet-50    & 14M ImageNet-21K      & 68,256,659 \\ \hline
			Instagram & ResNeXt-101  & 3.5B Instagram photos & 88,791,336 \\ \hline
		\end{tabular}
		}
		\label{pre-train models}
	\end{table}

	\begin{table*}
		\renewcommand\arraystretch{1.2}
		\caption{The effectiveness (p-value) of visual prompt IP verification on downstream datasets CIFAR-10, EuroSAT, and SVHN.}
		\resizebox{\linewidth}{!}{
		\begin{tabular}{c|c|ccc|ccc|ccc}
			\hline
			\multirow{2}{*}{Datasets} & \multirow{2}{*}{\begin{tabular}[c]{@{}c@{}}Prompt \\ attack\end{tabular}} & \multicolumn{3}{c|}{RLMVP}                                & \multicolumn{3}{c|}{FLMVP}                                & \multicolumn{3}{c}{ILMVP}                                 \\ \cline{3-11} 
			&                                                                           & RN50              & Instagram         & BiT-M             & RN50              & Instagram         & BiT-M             & RN50              & Instagram         & BiT-M             \\ \hline
			\multirow{5}{*}{CIFAR10}  & Unauthorized                                                              & 1.00E+00          & 1.00E+00          & 1.00E+00          & 1.00E+00          & 1.00E+00          & 1.00E+00          & 1.00E+00          & 1.00E+00          & 1.00E+00          \\
			& Fine-tuning                                                               & 1.00E+00          & 1.00E+00          & 1.00E+00          & 7.03E-02          & 1.06E-02          & 1.00E+00          & 7.02E-02          & 1.28E-02          & 1.00E+00          \\
			& Pruning-all                                                               & 1.00E+00          & 1.00E+00          & 1.00E+00          & 6.81E-03          & 5.85E-02          & 1.00E+00          & 3.96E-01          & 7.46E-04          & 3.20E-01          \\
			& Pruning-block                                                             & 1.00E+00          & 1.00E+00          & 1.00E+00          & 7.03E-02          & 1.95E-01          & 1.00E+00          & 3.96E-01          & 3.13E-02          & 1.00E+00          \\
			& Independent                                                               & \textbf{1.48E-24} & \textbf{6.45E-25} & \textbf{4.81E-33} & \textbf{2.04E-27} & \textbf{6.78E-29} & \textbf{3.16E-26} & \textbf{8.40E-23} & \textbf{1.68E-28} & \textbf{6.15E-32} \\ \hline
			\multirow{5}{*}{EuroSAT}  & Unauthorized                                                              & 1.00E+00          & 1.00E+00          & 1.00E+00          & 1.00E+00          & 1.00E+00          & 1.00E+00          & 1.00E+00          & 1.00E+00          & 1.00E+00          \\
			& Fine-tuning                                                               & 2.40E-01          & 3.84E-02          & 1.00E+00          & 3.20E-01          & 8.93E-02          & 3.20E-01          & 3.29E-03          & 1.00E+00          & 1.00E+00          \\
			& Pruning-all                                                               & 3.20E-01          & 5.66E-01          & 3.20E-01          & 4.93E-01          & 5.80E-01          & 1.91E-01          & 4.89E-04          & 1.77E-04          & 9.35E-03          \\
			& Pruning-block                                                             & 1.00E+00          & 1.00E+00          & 3.20E-01          & 8.20E-01          & 3.20E-01          & 3.20E-01          & 4.89E-04          & 2.40E-07          & 9.35E-03          \\
			& Independent                                                               & \textbf{1.60E-20} & \textbf{1.25E-21} & \textbf{7.88E-28} & \textbf{4.88E-19} & \textbf{9.19E-21} & \textbf{3.18E-21} & \textbf{3.65E-18} & \textbf{3.04E-22} & \textbf{9.75E-24} \\ \hline
			\multirow{5}{*}{SVHN}     & Unauthorized                                                              & 1.00E+00          & 1.00E+00          & 1.00E+00          & 1.00E+00          & 1.00E+00          & 1.00E+00          & 1.00E+00          & 1.00E+00          & 1.00E+00          \\
			& Fine-tuning                                                               & 1.81E-07          & 2.98E-03          & 3.20E-01          & 1.81E-01          & 2.83E-02          & 1.00E+00          & 9.58E-02          & 1.00E+00          & 1.00E+00          \\
			& Pruning-all                                                               & 8.40E-02          & 7.23E-07          & 1.00E+00          & 3.20E-01          & 1.58E-01          & 1.00E+00          & 8.32E-02          & 1.00E+00          & 1.00E+00          \\
			& Pruning-block                                                             & 8.40E-02          & 2.34E-05          & 1.00E+00          & 3.20E-01          & 3.20E-01          & 1.00E+00          & 8.32E-02          & 1.00E+00          & 1.00E+00          \\
			& Independent                                                               & \textbf{1.28E-21} & \textbf{1.81E-20} & \textbf{1.05E-25} & \textbf{3.11E-21} & \textbf{7.73E-25} & \textbf{3.62E-20} & \textbf{9.04E-18} & \textbf{1.48E-19} & \textbf{3.97E-22} \\ \hline
		\end{tabular}
	}
	\label{Differentiability}
	\end{table*}

	\section{Experiment}
	\subsection{Detailed Experimental Setups}
	
	\textbf{Datasets and pre-trained Models}. We conducted experiments on three benchmark visual image classification datasets: CIFAR10, EuroSAT, and SVHN. CIFAR10 comprises 60,000 images across ten object classes, EuroSAT includes 27,000 images spanning ten land use categories, and SVHN comprises 99,289 images representing 10 street view house numbers. The dataset was divided into non-overlapping train, test, and validation sets, with equal-sized test and validation sets. The validation set primarily assessed prompt robustness to various post-processing operations. Detailed partition information is provided in Table \ref{Dataset summary}.

	Three representative visual pre-training models were selected: ResNet trained on ImageNet-1K (RN50) \cite{he2016deep}, Big Transfer (BiT-M) \cite{kolesnikov2020big}, and ResNeXt trained on 3.5B Instagram images (Instagram) \cite{mahajan2018exploring}. Table \ref{pre-train models} describes the details of the pre-trained models.

	\textbf{Prompt Post-processing}. Attackers often employ post-processing techniques to circumvent IP detection mechanisms to modify prompt parameters, such as prompt fine-tuning and pruning. The goal is to ensure that visual prompts still pass WVPrompt validation even after these modifications. 
	
	\begin{figure*}[t]
		\centering
		\includegraphics[width=0.8\textwidth]{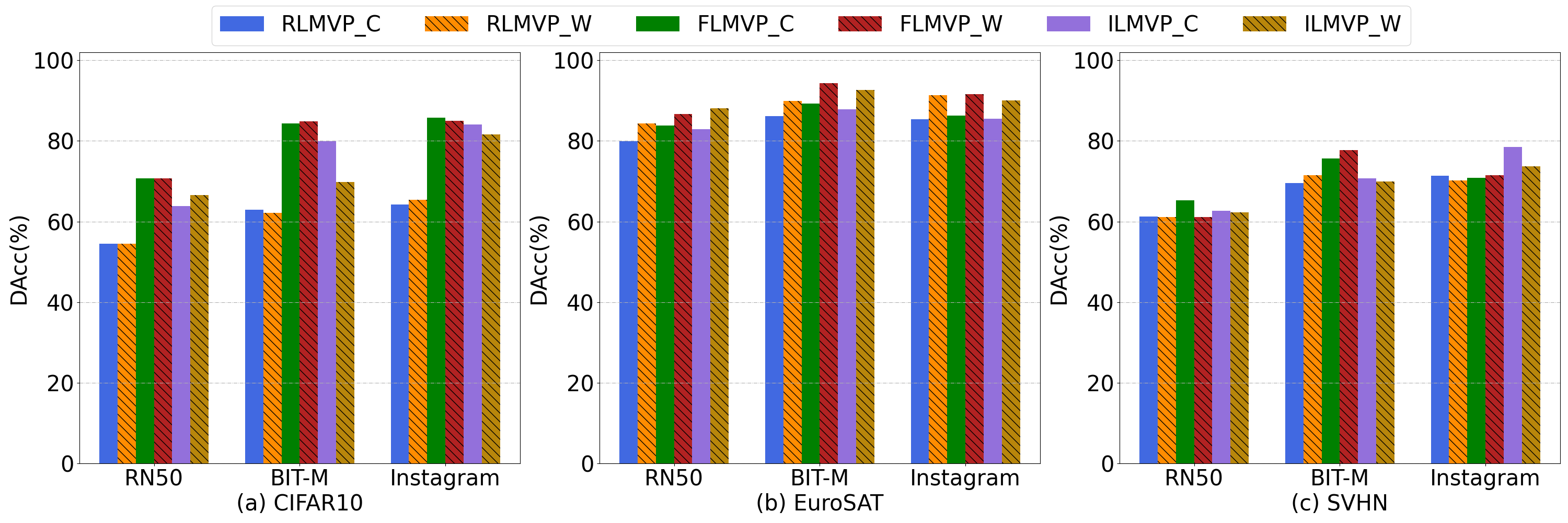} 
		\caption{Downstream accuracy of pre-trained large models indicated by clean and watermarked visual prompts. Here, RLMVP\_C and RLMVP\_W indicate clean visual prompts and watermarked visual prompts, respectively}
		\label{Harmlessness}
	\end{figure*}
	
	\begin{itemize}
	
	\item \textbf{Prompt fine-tuning:} This process involves updating all parameters in visual prompts. In this setting, we freeze the pre-trained model and conduct ten epochs of iterative training using the validation set to complete fine-tuning.
	\item \textbf{Prompt Pruning:} Pruning is a popular neural network model compression method and has been widely used in previous work to remove embedded identity watermark information. Similarly, this paper proposes a method of prompt pruning, which is primarily implemented by iteratively resetting the $q\%$ parameters with the smallest absolute value of the visual prompt to 0. In this case, we gradually increase $q$ from 0.1 to 0.9 in steps of 0.1. Generally speaking, padding-based prompts consist of four blocks (or layers): top, bottom, left, and right, and the parameter sizes of different blocks overlap. Therefore, this paper proposes two pruning methods: independent pruning of each block (pruning-block) and joint pruning of all blocks (pruning-all).
	\end{itemize}
	
	\begin{figure}[t]
		\centering
		\includegraphics[width=0.5\textwidth]{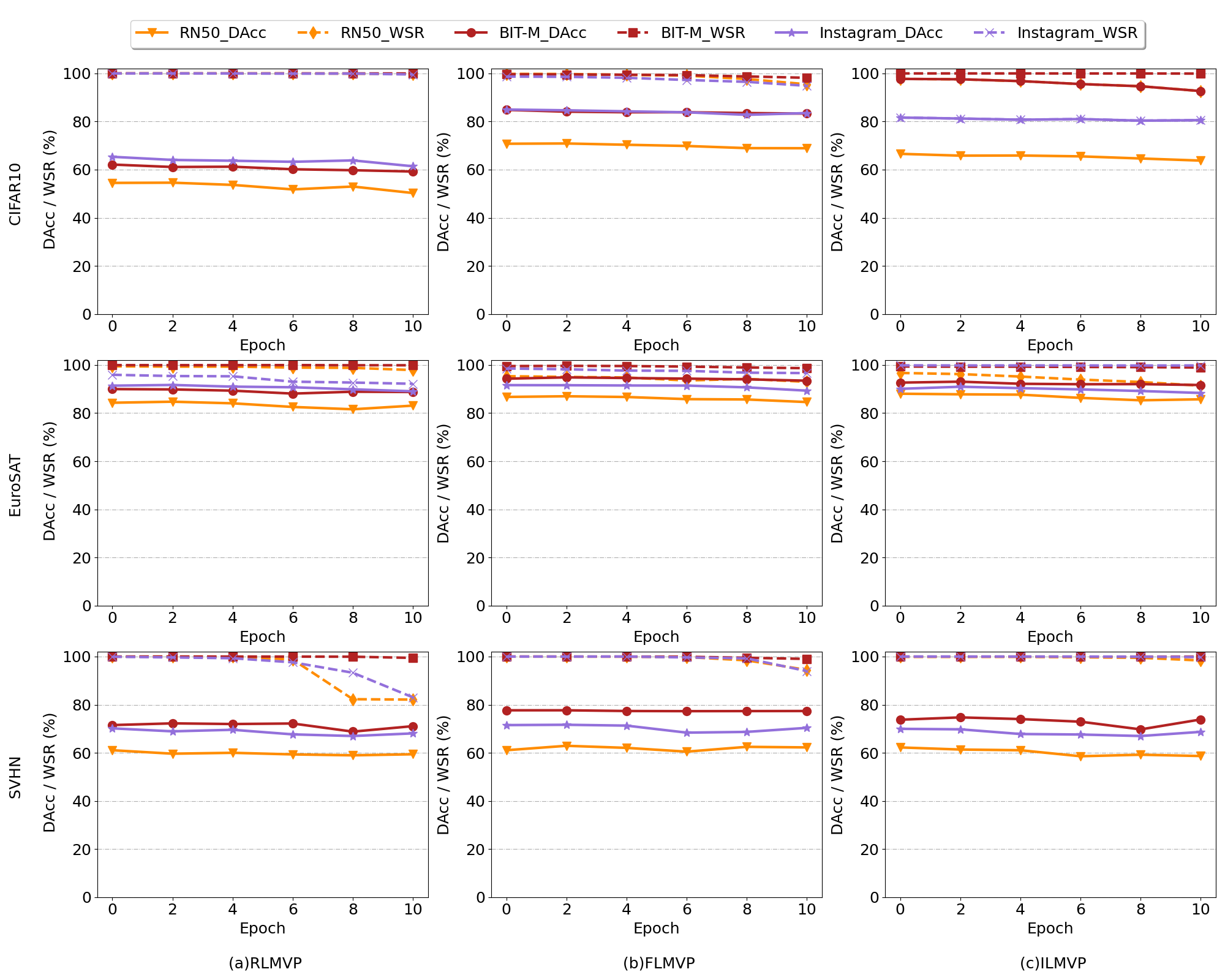} 
		\caption{The accuracy of downstream task and watermarking success rate are obtained under different fine-tuning epochs}
		\label{fineturn}
	\end{figure}
	
	\begin{figure}[t]
		\centering
		\includegraphics[width=0.5\textwidth]{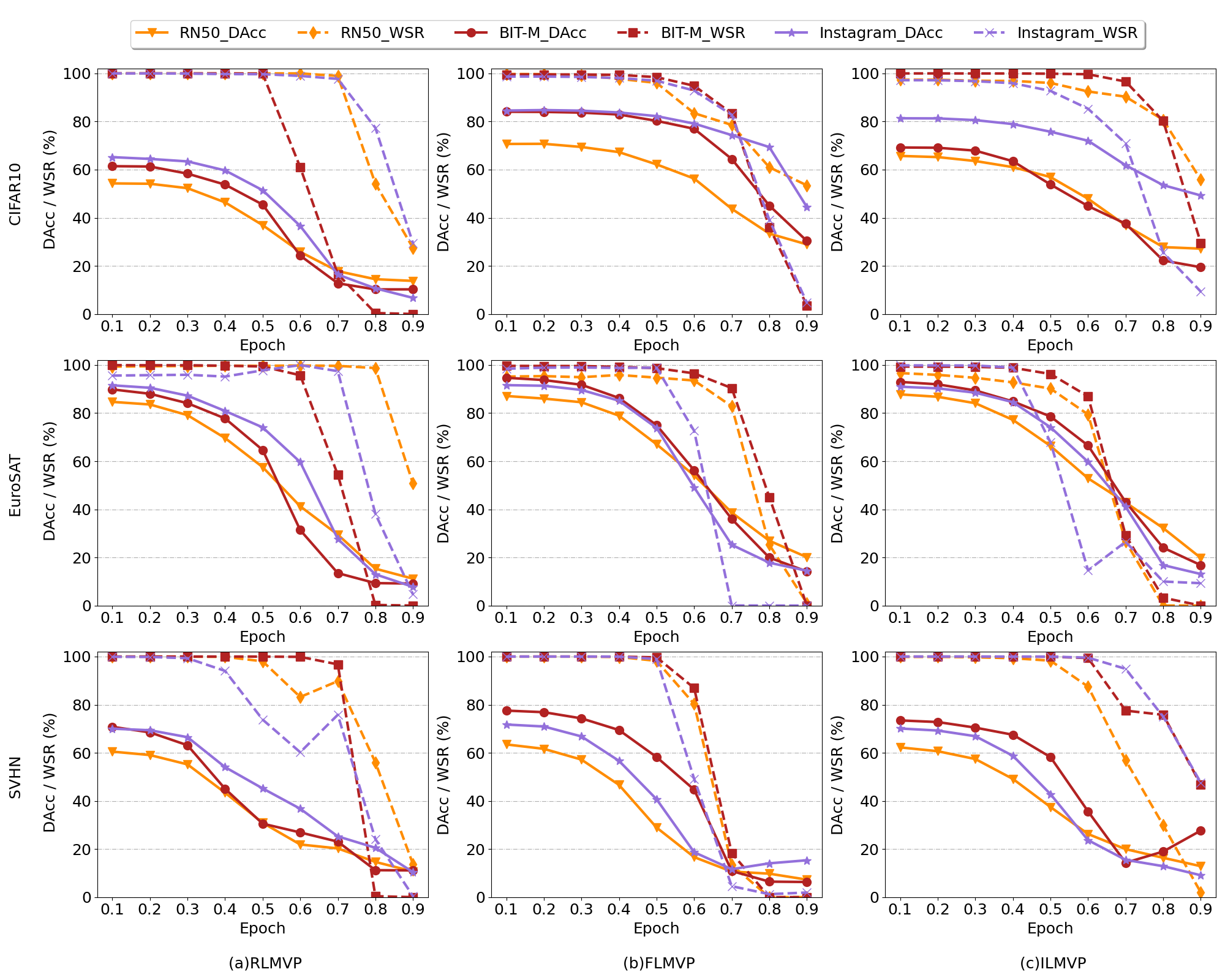} 
		\caption{The accuracy of downstream task and watermarking success rate in the prompt block pruning}
		\label{purn_block}
	\end{figure}
	
	\begin{figure}[t]
		\centering
		\includegraphics[width=0.5\textwidth]{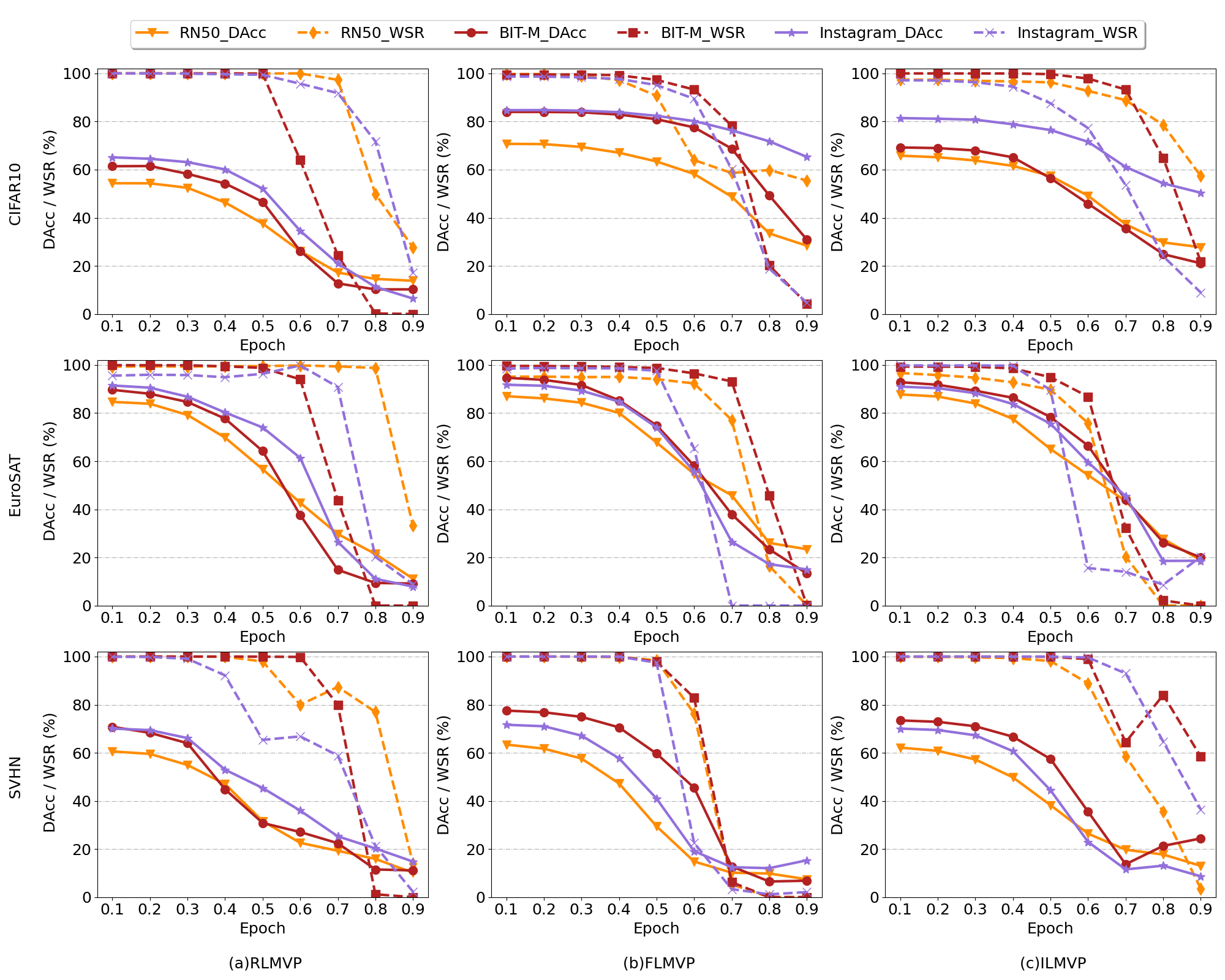} 
		\caption{The accuracy and watermarking success rate of downstream tasks in all prompts for pruning}
		\label{purn_all}
	\end{figure}
	
	\textbf{Evaluation Metrics:} To verify the effectiveness of WVPrompt in watermark verification, we conducted a two-sample hypothesis test and used p-values to evaluate our method (Proposition 1). Additionally, we utilized two metrics to evaluate the effect of watermark injection: downstream task accuracy (DAcc) and watermark success rate (WSR). DAcc represents the accuracy of the watermarked prompts combined with the pre-trained model on the clean test set, which is used to verify the harmlessness of WVPrompt. WSR represents the accuracy of the test set with backdoor triggers, which is used to prove the robustness of WVPrompt. Generally, a more minor impact on DAcc and a larger WSR imply greater watermark effectiveness.

	\textbf{Detailed settings:} We followed the default experimental settings in visual prompt learning \cite{bahng2022exploring}. The prompt template was padded with a size $p$ of 30 on all sides. The number of parameters per prompt is $2 \times C \times p \times (H+ W- 2p)$, where $p, C, H$, and $W$ are the prompt size, image channels, height, and width, respectively. All images were resized to 224 × 224 to match the input to the pre-trained model. Therefore, the number of parameters per prompt is 69,840. In the specific downstream task, 100 epochs were used, with a batch size of 128, an initial learning rate of 40, a momentum of 0.9, and a stochastic gradient descent (SGD) optimizer with a cross-entropy loss function. During prompt watermark injection, a 4*4 black and white block was used as the backdoor trigger, added to the lower right corner of the sample image, and $\beta$ was set to 1. The first category in each public dataset served as the target category. We adopt the first category in each public dataset as the target category $y_{t}$, such as "automobile" for CIFAR10, "1" for SVHN, and "forest" for EuroSAT. Unless otherwise specified, the poisoning rate $r_{p}$ is 0.1.

	\subsection{Effectiveness} 
	
	In this section, we present experiments to evaluate the effectiveness of WVPrompt under three representative visual prompt learning methods (i.e., RLMVP, FLMVP, ILMVP). Specifically, we randomly select m as 100 samples and input them into the poisoner G to obtain samples with triggers. These samples are combined with the watermarked victim prompts and suspicious prompts before input into the large-scale pre-training model. We obtain two sets of label sequences, P1 and P2, output by the pre-trained model. Finally, we use a T hypothesis test to calculate its p-value.

	To avoid the IP detection mechanism, attackers often use the post-processing technology described in Section 4.1.2 to modify stolen prompt parameters inexpensively. Therefore, this study evaluates p-values under five prompt design scenarios: unauthorized, fine-tuning, pruning-blocks, pruning-all, and independent prompts. Among these scenarios, in the first four, the suspicious prompt is considered pirated and infringing on the IP of the victim VPaaS developer. The experimental results are depicted in Table \ref{Differentiability}, demonstrating that our method accurately identifies piracy with high confidence (i.e., $p>10^{-7} $), with $96\%$ having $p>10^{-4} $. There is minimal evidence to reject the null hypothesis $H_{0}$. Conversely, there is a smaller $p$ value ($p<10^{-19} $) for independent prompts, indicating that the null hypothesis $H_{0}$ can be accepted with strong confidence. Thus, WVPrompt effectively verifies the ownership of visual prompts.

	\subsection{Harmlessness}
	One of the primary goals of WVPrompt was to maintain its usefulness in the original downstream tasks. To verify the harmlessness of prompt watermark injection, we first compared the accuracy of clean and watermarked prompts on the clean test set of downstream tasks. The results are shown in Figure \ref{Harmlessness}, where $xx_C$ and $xx_W$ represent the clean and watermarked prompts optimized under the $xx$ prompt learning method. In most datasets and pre-trained model structures, the decrease in DAcc of watermarked prompts is less than $5\%$, indicating that the injected watermark minimally affects prompt performance. Moreover, the DAcc of some watermarked prompts is even higher than that of clean prompts. This could be attributed to the training set for watermarked prompts, which consists of the clean, prompt training set $D_{c}$ and the toxic sample set $D_{p}$, enhancing the generalization capabilities of watermarked prompts. Additionally, in approximately 90\% of experiments, FLMVP exhibits superior performance compared to ILMVP and RLMVP. This may be because FLMVP selects the label with the highest output probability calculated on the downstream task dataset. In contrast, ILMVP gives more consideration to the interpretability of label mappings, and RLMVP adopts a simple one-to-one mapping of the top K model labels. Overall, the WVPrompt method demonstrates harmlessness.

	\begin{table}
		\renewcommand\arraystretch{1.2}
		\caption{Influence of different poisoning rates on downstream task accuracy (DAcc) and watermarking success rate (WAR) under three vision prompt learning methods}
		\resizebox{\linewidth}{!}{
		\begin{tabular}{c|c|c|ccccc}
			\hline
			\multirow{2}{*}{Prompt} & \multirow{2}{*}{Datasets} & \multirow{2}{*}{Metrics} & \multicolumn{5}{c}{Posion rate}       \\ \cline{4-8} 
			&                           &                          & 0.01  & 0.03  & 0.05  & 0.1   & 0.15  \\ \hline
			\multirow{6}{*}{RLMVP}  & \multirow{2}{*}{CIFAR10}  & DAcc                     & 61.85 & 61.93 & 62.12 & 61.74 & 61.97 \\
			&                           & WSR                      & 99.94 & 100   & 100   & 100   & 100   \\ \cline{2-8} 
			& \multirow{2}{*}{EuroSAT}  & DAcc                     & 89.93 & 90.22 & 89.98 & 89.61 & 89.94 \\
			&                           & WSR                      & 71.83 & 99.85 & 99.94 & 99.98 & 100   \\ \cline{2-8} 
			& \multirow{2}{*}{SVHN}     & DAcc                     & 70.53 & 71.19 & 71.52 & 70.93 & 63.48 \\
			&                           & WSR                      & 99.97 & 100   & 100   & 100   & 99.99 \\ \hline
			\multirow{6}{*}{FLMVP}  & \multirow{2}{*}{CIFAR10}  & DAcc                     & 84.7  & 84.88 & 84.84 & 84.93 & 84.51 \\
			&                           & WSR                      & 93.96 & 99.46 & 99.63 & 99.47 & 99.77 \\ \cline{2-8} 
			& \multirow{2}{*}{EuroSAT}  & DAcc                     & 94.24 & 94.31 & 94.3  & 94.81 & 94.07 \\
			&                           & WSR                      & 87.09 & 97.89 & 99.54 & 99.98 & 99.96 \\ \cline{2-8} 
			& \multirow{2}{*}{SVHN}     & DAcc                     & 76.82 & 76.87 & 77.67 & 77.24 & 76.36 \\
			&                           & WSR                      & 99.85 & 99.99 & 100   & 100   & 100   \\ \hline
			\multirow{6}{*}{ILMVP}  & \multirow{2}{*}{CIFAR10}  & DAcc                     & 73.25 & 72.32 & 69.77 & 70.03 & 70.25 \\
			&                           & WSR                      & 99.16 & 99.56 & 100   & 100   & 100   \\ \cline{2-8} 
			& \multirow{2}{*}{EuroSAT}  & DAcc                     & 93.07 & 92.74 & 92.67 & 92.69 & 92.46 \\
			&                           & WSR                      & 92.54 & 98.48 & 99.3  & 100   & 99.91 \\ \cline{2-8} 
			& \multirow{2}{*}{SVHN}     & DAcc                     & 73.85 & 73.88 & 73.75 & 73.17 & 74.39 \\
			&                           & WSR                      & 97.44 & 99.92 & 99.99 & 100   & 100   \\ \hline
		\end{tabular}
			}
		\label{Poison rate}
	\end{table}
	
	\subsection{Robustness}
	To evade IP protection mechanisms, attackers may modify the parameters of stolen visual prompts through prompt fine-tuning and pruning. Therefore, assessing the robustness of watermarked prompts under these post-processing techniques is crucial.

	\textbf{prompt fine-tuning.} We fine-tune all parameters in the watermarked visual prompts using a clean validation set of the same size as the test set. Other experimental parameters remained constant. Figure \ref{fineturn} shows the changes in DAcc and WSR when fine-tuned on CIFAR10, EuroSAT, and SVHN datasets. It is observed that as the number of epochs increases, WSR remains unchanged or decreases slightly. The most significant drop occurs after eight fine-tuning rounds on the SVHN data in RMLVP prompt learning mode. However, overall, it remains above $82\%$. Meanwhile, the change amplitude of DAcc remains within $3\%$ in $92\%$ of cases, which we consider to be a normal change in visual prompt learning.

	\textbf{Prompt pruning.} We set the first $q\%$ (from $10\%$ to $90\%$) minimum absolute value parameters in the prompt parameters to 0 in scenarios by pruning-block or pruning-all. Finally, the performance of prompt pruning is measured using the downstream task test set. Figures \ref{purn_block} and \ref{purn_all} illustrate that WSR and DAcc gradually decrease with increasing pruning rates. Overall, pruning-all is more stable than pruning-block. Specifically, when deleting parameters below $40\%$ by pruning-block and $50\%$ for pruning-all, WSR, and Dacc are not significantly affected. However, from $70\%$ to $90\%$, both WSR and Dacc decrease significantly. This decrease is likely due to the fact that compared to the pre-trained model, there are too few parameters in the prompt, and each bit plays a larger role.

	\subsection{Ablation experiment}
	In this section, we quantitatively analyze the core parameters in WVPrompt, including the impact of the sample number $m$ and the poisoning rate $r_{p}$ in the T-test. For simplicity, we discuss different visual prompt learning methods and different datasets under the pre-training model of BiT\_M.

	\begin{table*}
		\renewcommand\arraystretch{1.2}
		\caption{The verification effectiveness (p-value) of our WVPrompt with different sampling numbers on pre-trained model BIT-M, where UNA represents an unauthorized prompt, and IND is an independent prompt}
		\resizebox{\linewidth}{!}{
		\begin{tabular}{c|c|c|ccccccc}
			\hline
			\multirow{2}{*}{Prompt} & \multirow{2}{*}{Datasets} & \multirow{2}{*}{\begin{tabular}[c]{@{}c@{}}Prompt \\ attack\end{tabular}} & \multicolumn{7}{c}{Sampling numbers}                                                     \\ \cline{4-10} 
			&                           &                                                                           & 20         & 40         & 60         & 80         & 100        & 120        & 140        \\ \hline\multirow{6}{*}{RMLVP}  & \multirow{2}{*}{CIFAR10}  & UNA                                                                       & 1          & 1          & 1          & 1          & 1          & 1          & 1          \\
			&                           & IND                                                                       & 5.5185E-07 & 1.8753E-14 & 3.4248E-20 & 5.6053E-27 & 4.8129E-33 & 2.3140E-36 & 2.1300E-41 \\ \cline{2-10} 
			& \multirow{2}{*}{EuroSAT}  & UNA                                                                       & 1          & 1          & 1          & 1          & 1          & 1          & 1          \\
			&                           & IND                                                                       & 4.8151E-06 & 1.5886E-09 & 4.6460E-14 & 1.4870E-19 & 6.1677E-25 & 1.6033E-28 & 2.2508E-33 \\ \cline{2-10} 
			& \multirow{2}{*}{SVHN}     & UNA                                                                       & 1          & 1          & 1          & 1          & 1          & 1          & 1          \\
			&                           & IND                                                                       & 5.4031E-06 & 1.5056E-11 & 3.3381E-16 & 3.4689E-21 & 1.0506E-25 & 2.2509E-30 & 3.7486E-36 \\ \hline\multirow{6}{*}{FLMVP}  & \multirow{2}{*}{CIFAR10}  & UNA                                                                       & 1          & 1          & 1          & 1          & 1          & 1          & 1          \\
			&                           & IND                                                                       & 4.2185E-06 & 3.0027E-11 & 2.6334E-16 & 1.3899E-22 & 3.1631E-26 & 1.1699E-29 & 2.4198E-35 \\ \cline{2-10} 
			& \multirow{2}{*}{EuroSAT}  & UNA                                                                       & 1          & 1          & 1          & 1          & 1          & 1          & 1          \\
			&                           & IND                                                                       & 2.0079E-05 & 5.3114E-09 & 1.2874E-13 & 1.8252E-17 & 4.6647E-22 & 2.3583E-25 & 6.5361E-30 \\ \cline{2-10} 
			& \multirow{2}{*}{SVHN}     & UNA                                                                       & 1          & 1          & 1          & 1          & 1          & 1          & 1          \\
			&                           & IND                                                                       & 9.0399E-05 & 1.6126E-09 & 2.4874E-14 & 9.4691E-17 & 3.6200E-20 & 6.1630E-23 & 2.2666E-27 \\ \hline\multirow{6}{*}{ILMVP}  & \multirow{2}{*}{CIFAR10}  & UNA                                                                       & 1          & 1          & 1          & 1          & 1          & 1          & 1          \\
			&                           & IND                                                                       & 9.2260E-07 & 7.9537E-14 & 4.3900E-21 & 9.1014E-28 & 6.1494E-32 & 4.0947E-35 & 1.2745E-39 \\ \cline{2-10} 
			& \multirow{2}{*}{EuroSAT}  & UNA                                                                       & 1          & 1          & 1          & 1          & 1          & 1          & 1          \\
			&                           & IND                                                                       & 1.2460E-05 & 2.8095E-08 & 6.5211E-13 & 6.7264E-17 & 4.1979E-21 & 3.7022E-25 & 5.7549E-30 \\ \cline{2-10} 
			& \multirow{2}{*}{SVHN}     & UNA                                                                       & 1          & 1          & 1          & 1          & 1          & 1          & 1          \\
			&                           & IND                                                                       & 1.8703E-04 & 1.4959E-09 & 3.1348E-14 & 3.2656E-17 & 3.9669E-22 & 1.2350E-25 & 8.4701E-30 \\ \hline
			
		\end{tabular}
		}
		
		\label{Sampling Number}
	\end{table*}

	\textbf{The influence of sample number $m$}. When conducting the T-test, triggers are added to $m$ clean test samples, which are input into the pre-training model along with the victim's watermarked and suspicious prompts. Two sets of output sequences are obtained, and their P-values are calculated. As shown in Table \ref{Sampling Number}, the verification performance improves with an increase in the sample number $m$. These results are expected as our approach achieves promising WSR. Generally, a larger $m$ reduces the adverse impact of randomness in verification, leading to a higher confidence level. However, it's essential to note that a larger $m$ requires more queries to the model API, which can be expensive and raise suspicion.

	\textbf{For the impact of the poisoning rate $r_{p}$}. We evaluated the changes in WSR and DAcc of watermarked prompts across poisoning rates ranging from $1\%$ to $15\%$. As shown in Table \ref{Poison rate}, WSR increases with the poisoning rate $r_{p}$ in all cases. These results indicate that defenders can enhance verification confidence using a relatively large $r_{p}$. Notably, even with a small poisoning rate (e.g., $1\%$), nearly all evaluated attacks achieve watermark success rates above $90\%$. However, in some scenarios, the downstream task accuracy decreases as $r_{p}$ increases. Specifically, for the RLMVP algorithm on the SVHN dataset, when $r_{p}$ is $15\%$, the DAcc value is reduced by about $7\%$ compared to when $r_{p}$ is $1\%$. In other words, there is a trade-off between WSR and DAcc to some extent. Defenders should allocate $r_{p}$ based on specific needs in practice.

	\section{Related Work}
	
	\textbf{Vision prompt learning.} Bahng et al. \cite{bahng2022exploring} first introduced the concept of visual prompt learning (VPL), drawing inspiration from context learning and prompt learning in natural language processing (NLP) \cite{li2021prefix,lester2021power,shin2020autoprompt}. Prior to this, a similar approach known as model reprogramming or adversarial reprogramming \cite{elsayed2018adversarial,neekhara2018adversarial,neekhara2022cross,zheng2023adversarial,zhang2022fairness} had been utilized within the field of computer vision. These methods mainly focus on learning common input patterns (e.g., pixel perturbations) and output label mapping (LM) functions, enabling pre-trained models to adapt to new downstream tasks without necessitating model fine-tuning. Another avenue of research is visual prompt tuning \cite{sohn2023visual,gao2022visual,wu2023approximated,jia2022visual}, where visual prompts come in the form of additional model parameters but are usually limited to visual transformers. In terms of specific applications, VPL has not only found traction in visual models but has also garnered attention in language models like CLIP \cite{radford2021learning}. Research \cite{bahng2022exploring} shows that with the assistance of CLIP, VPL can generate prompt patterns of image data without resorting to source-target label mapping. In \cite{khattak2023maple}, VPL and text prompts are jointly optimized in the CLIP model to obtain better performance. Additionally,  in domains characterized by data scarcity, such as biochemistry, VPL has been proven to be effective in cross-domain transfer learning \cite{chen2021adversarial,neekhara2022cross,yen2021study}. Despite these advancements, intellectual property protection concerning visual cue learning remains largely unexplored.
	
	\textbf{Watermark.} Watermarking is a concept traditionally used in media such as audio and video \cite{saini2014survey}, and is characterized by hiding information in data and remaining imperceptible to identify the authenticity or attribution of the data. In recent years, watermarking technology has expanded into the field of intellectual property protection for machine learning models \cite{adi2018turning,lukas2022sok,li2023black}. Watermarking methods of deep neural networks (DNN) are mainly divided into two categories: white-box and black-box watermarking. Black-box  watermarks \cite{shafieinejad2021robustness,zhang2018protecting} are easier to verify than white-box watermarks \cite{uchida2017embedding,li2021survey} because in the former verification only the stolen model access service is used to verify the ownership of the deep learning model, while the latter requires the model owner to access all parameters of the model to extract the watermark. Furthermore, black-box watermarking is superior to white-box watermarking as it is more likely to be resilient to statistical attacks. However, to the best of our knowledge, there is a lack of relevant research on utilizing watermarking technology for prompt IP protection in both black-box and white-box environments. Only the recent PromptCARE \cite{yao2023promptcare} has begun to study prompt watermarking in the field of natural language in black-box scenarios. However, due to the natural difference between textual language and visual images, this watermarking technology cannot be directly applied to visual prompt IP protection.

	\section{Conclusion}
	Visual prompts effectively address computational and storage challenges when deploying large models across diverse downstream tasks, making them a valuable asset for developers. However, no technology is currently available to protect the ownership of visual prompts. Reducing prompt parameters and various post-processing techniques pose challenges in effectively embedding and preserving identity watermarking information. We introduce a black-box watermarking method called WVPrompt, comprising two stages: watermark injection and validation. Specifically, it embeds the model's watermark into the visual prompt by poisoning the dataset and efficiently validates WVPrompt's effectiveness using hypothesis testing. Experimental results demonstrate the superiority of WVPrompt.
	
	\bibliographystyle{unsrt}
	\bibliography{reference}
	
\end{document}